\RequirePackage{fix-cm}
\documentclass{article}

\usepackage{arxiv}

\usepackage{lineno,hyperref}
\modulolinenumbers[5]

\usepackage[utf8]{inputenc} % allow utf-8 input
\usepackage[T1]{fontenc}    % use 8-bit T1 fonts
\usepackage{hyperref}       % hyperlinks
\usepackage{url}            % simple URL typesetting
\usepackage{booktabs}       % professional-quality tables
\usepackage{amsfonts}       % blackboard math symbols
\usepackage{amsmath}        % maths formatting
\usepackage{nicefrac}       % compact symbols for 1/2, etc.
\usepackage{microtype}      % microtypography
\usepackage{lipsum}
\usepackage[linesnumbered,ruled,vlined]{algorithm2e}

\usepackage{graphicx}
\usepackage{caption}
\usepackage{subcaption}
\usepackage{textcomp}
\usepackage[dvipsnames]{xcolor}
\usepackage{siunitx}
\usepackage[sorting=none]{biblatex}
\graphicspath{ {graphics/} }

\usepackage{etoolbox,refcount}
\usepackage{multicol}
\newcounter{countitems}
\newcounter{nextitemizecount}
\newcommand{\setupcountitems}{%
  \stepcounter{nextitemizecount}%
  \setcounter{countitems}{0}%
  \preto\item{\stepcounter{countitems}}%
}
\makeatletter
\newcommand{\computecountitems}{%
  \edef\@currentlabel{\number\c@countitems}%
  \label{countitems@\number\numexpr\value{nextitemizecount}-1\relax}%
}
\newcommand{\nextitemizecount}{%
  \getrefnumber{countitems@\number\c@nextitemizecount}%
}
\newcommand{\previtemizecount}{%
  \getrefnumber{countitems@\number\numexpr\value{nextitemizecount}-1\relax}%
}
\makeatother
\newenvironment{AutoMultiColItemize}{%
\ifnumcomp{\nextitemizecount}{>}{3}{\begin{multicols}{2}}{}%
\setupcountitems\begin{itemize}}%
{\end{itemize}%
\unskip\computecountitems\ifnumcomp{\previtemizecount}{>}{3}{\end{multicols}}{}}

\newcommand{\norm}[1]{\left\lVert#1\right\rVert}
\DeclareMathOperator{\Tr}{Tr}

\begin{document}
\title{Evaluating the Clinical Realism of Synthetic Chest X-Rays Generated Using Progressively Growing GANs}

\author{
  Bradley Segal\\
  Biomedical Engineering Research Group\\
  School of Electrical and Information Engineering\\
  University of the Witwatersrand, Johannesburg\\
  South Africa \\
  \texttt{Brad@Segal.co.za} \\

   \And
  David M. Rubin \\
  Biomedical Engineering Research Group\\
  School of Electrical and Information Engineering\\
  University of the Witwatersrand, Johannesburg\\
  South Africa \\
  \texttt{David.Rubin@wits.ac.za} \\
  
   \And
  Grace Rubin \\
  Helen Joseph Hospital\\
  Department of Radiation Sciences\\
  Division of Radiology\\
  University of the Witwatersrand, Johannesburg\\
  South Africa \\
  \texttt{rubingrace@gmail.com} \\
  
  \And
  Adam Pantanowitz \\
  Biomedical Engineering Research Group\\
  School of Electrical and Information Engineering\\
  University of the Witwatersrand, Johannesburg\\
  South Africa \\
  \texttt{Adam.Pantanowitz@wits.ac.za} \\
}

\maketitle

\begin{abstract}
\label{sec:abstract}
Chest x-rays are a vital diagnostic tool in the workup of many patients. Similar to most medical imaging modalities, they are profoundly multi-modal and are capable of visualising a variety of combinations of conditions. There is an ever pressing need for greater quantities of labelled images to drive forward the development of diagnostic tools, however this is in direct opposition to concerns regarding patient confidentiality which constrains access through permission requests and ethics approvals. Previous work has sought to address these concerns by creating class-specific generative adversarial networks (GANs) that synthesise images to augment training data. These approaches cannot be scaled as they introduce computational trade offs between model size and class number which places fixed limits on the quality that such generates can achieve. We address this concern by introducing latent class optimisation which enables efficient, multi-modal sampling from a GAN and with which we synthesise a large archive of labelled generates. We apply a Progressive Growing GAN (PGGAN) to the task of unsupervised x-ray synthesis and have radiologists evaluate the clinical realism of the resultant samples. We provide an in depth review of the properties of varying pathologies seen on generates as well as an overview of the extent of disease diversity captured by the model. We validate the application of the Fréchet Inception Distance (FID) to measure the quality of x-ray generates and find that they are similar to other high resolution tasks. We quantify x-ray clinical realism by asking radiologists to distinguish between real and fake scans and find that generates are more likely to be classed as real than by chance, but there is still progress required to achieve true realism. We confirm these findings by evaluating synthetic classification model performance on real scans. We conclude by discussing the limitations of PGGAN generates and how to achieve controllable, realistic generates going forward. 

We release our source code, model weights, and an archive of labelled generates.

\end{abstract}

\section{Introduction}
\label{sec:introduction}
The chest radiograph (CXR) is the most common diagnostic radiological procedure \cite{adamCurrentStatusThoracic2021} and is commonly used to screen, diagnose or monitor conditions across a variety of clinical contexts. This ubiquity has resulted in substantial research interest in the automated diagnosis of such films~\cite{kermanyIdentifyingMedicalDiagnoses2018,majkowskaChestRadiographInterpretation2020,rajpurkarCheXNetRadiologistLevelPneumonia2017}, with a recent surge in activity due to the COVID-19 pandemic~\cite{ozturkAutomatedDetectionCOVID192020,shamoutArtificialIntelligenceSystem2020,wangCOVIDNetTailoredDeep2020}. State-of-the-art (SOTA) models are capable of radiologist-level performance across a subset of pathologies in only a fraction of the time needed for human review~\cite{irvinCheXpertLargeChest2019,rajpurkarDeepLearningChest2018,rajpurkarCheXNetRadiologistLevelPneumonia2017}. This activity in CXRs represents a section of the development of Computer Aided Diagnostic (CAD) systems that aim to provide tangible benefit to clinicians and patients alike by reducing diagnostic turnaround times, minimising errors, and supporting the clinical decision making process~\cite{rajpurkarDeepLearningChest2018,santosArtificialIntelligenceMachine2019}. Large scale, anonymised, public imaging datasets underscore these efforts by providing the necessary clinical data for the training of CAD models~\cite{irvinCheXpertLargeChest2019,johnsonMIMICCXRDeidentifiedPublicly2019,wangChestXray8HospitalscaleChest}. The development of these archives is both time consuming and costly, requiring extensive expert labelling and anonymisation of patient protected information prior to release. Radiology reports produced during clinical practice are typically used as surrogates for expert review and are mined for diagnostic labels~\cite{irvinCheXpertLargeChest2019,rajpurkarDeepLearningChest2018,rajpurkarCheXNetRadiologistLevelPneumonia2017}. Anonymisation of images involves the detection of annotations containing protected patient information such as names that may be contained within the pixel data, or the removal of metadata that may be contained within digital imaging and communications in medicine (DICOM) files~\cite{johnsonMIMICCXRDeidentifiedPublicly2019}. This process does not adjust the image data of a particular scan, which may result in patient re-identification if visual elements of the image are rare. 

Medical imagery exhibits multi-modal long-tailed distributions with significant heterogeneity in the presentation of similar disease patterns due to equipment, scan technique, and patient
variability~\cite{zhouReviewDeepLearning2020}. This issue is demonstrated by most large CXR archives as they are limited to a single clinical site which reduces the captured variability of the factors of heterogeneity~\cite{zhouReviewDeepLearning2020}. The resulting sparsity of training data has obvious implications for making the development of robust diagnostic models more challenging. In addition, alterations in protected patient information such as sex, age, ethnicity, and socioeconomic class have been demonstrated to produce biases in diagnostic performance not significantly explained by variations in disease prevalence with improvements only seen with usage of multi-source image datasets~\cite{seyyed-kalantariCheXclusionFairnessGaps2020}. 

There is a clear need for methodologies capable of resolving this data disparity while retaining or even augmenting privacy. Generative adversarial networks (GANs) have received significant attention as a potential solution for their ability to synthesise medical images~\cite{beersHighresolutionMedicalImage2018,salehinejadSynthesizingChestXRay2019a,togoSyntheticGastritisImage2019} without compromising patient confidentiality as they learn to replicate the source distribution without access to the underlying training data~\cite{yiGenerativeAdversarialNetwork2019}. GANs capable of high fidelity image synthesis would address data challenges in machine learning and in medical education and training by:
\begin{itemize}
\item Creating datasets that do not compromise patient confidentiality, but provide the same diagnostic outcomes;
\item Providing class-balanced datasets; 
\item Generating augmented images by semantic variation in visual content without altering the diagnosis; and
\item Synthesising images with specified target pathologies
\end{itemize}

This study serves to comprehensively evaluate the medical plausibility of synthetic CXRs as well as their applications to diagnostic radiology~\footnote{An ethics waiver was issued for this work by the Human Research Ethics Committee (Medical) of the University of the Witwatersrand, Johannesburg on 11/08/2020.}. We evaluate the Progressively Growing GAN (PGGAN)~\cite{karrasProgressiveGrowingGANs2018} methodology applied to generate multi-modal, megapixel resolution CXRs. We provide domain expert examination of the properties of generated images as well as a review of expert discrimination between real and generated samples. In addition, we provide the following contributions:
\begin{itemize}
\item An evaluation of the applicability of the Fréchet Inception Distance (FID) for automatically evaluating x-ray generates.
\item A proposal for extracting images for a specific pathology through a modified latent space search.
\item A proposal for generating patient image series through local pathology sampling.  
\item An evaluation of the performance of synthetic image datasets derived from multi-modal generators.
\end{itemize}

Source code is available~\cite{segal2021CXRPGGANCode}, as well as model weights and a collection of labelled generated images~\cite{segal2021CXRPGGANModel}.

\section{Background}

\subsection{Generative Adversarial Networks}
GANs are a form of implicit generative model rooted in game theory that learns to reproduce an unseen training distribution through competitive optimisation. The prototypical GAN consists of a pair of neural networks in opposition to one another: a generator~($G$) whose goal is to generate plausible samples, and a discriminator~($D$) whose goal is to distinguish such samples from real samples~($x$) drawn from a training distribution~\cite{goodfellowGenerativeAdversarialNetworks2014}. The generator learns to sample from a latent vector~($z$) to produce a sample~($G(z)$) of the generated~($\mathbb P_{g}$) distribution which is similar to a chosen reference distribution~($\mathbb P_{r}$). The training signal is provided by how effectively such a generate is classified as real by the discriminator~($D(G(z))$). This configuration is optimal when the generator produces samples indistinguishable from the reference set and the discriminator can no longer learn to detect generates~\cite{meschederWhichTrainingMethods2018,salimansImprovedTechniquesTraining2016}. The configuration can be interpreted as a two-player minimax game with the following value function:

\begin{equation} \label{eq:1}
\min_G \max_D V(D,G) = \underset{x \sim \mathbb P_{r}}{\mathbb E}[\log D(x)] + \underset{x \sim \mathbb P_{g}}{\mathbb E}[\log(1 - D(G(x)))]
\end{equation}
wherein the discriminator ($D$) attempts to maximise the value function by separating real from generated images and the generator ($G$) attempts to minimise it by producing samples that cannot be detected by the discriminator.

The GAN formulation has found utilisation for a variety of tasks, namely image-to-image translation~\cite{isolaImagetoImageTranslationConditional2018,zhuUnpairedImagetoImageTranslation2018}, image super-resolution~\cite{ledigPhotoRealisticSingleImage2017,youCTSuperresolutionGAN2020}, and semantic image editing~\cite{shenInterpretingLatentSpace2020} to name just a few. Perhaps the most common usage is that of image synthesis, wherein a GAN attempts to produce images that mirror a reference image set. The Deep Convolutional GAN~\cite{radfordUnsupervisedRepresentationLearning2016} was an initial step towards improving resolutions by making use of convolutional layers as opposed to the fully connected approach of the original GAN. 

This however, did not resolve fundamental problems with training instability that caused models to fail to converge or resulted in mode dropping, a phenomenon where image variability is sacrificed for image quality and at worst can result in a the generation of only a handful of images~\cite{arjovskyPrincipledMethodsTraining2017}. These issues become more prominent as image resolution increases as the discriminator is able to more easily detect generated samples~\cite{karrasProgressiveGrowingGANs2018}. This places significant constraints on the upper limit of resolutions that a DCGAN model can achieve.

Recent work has adapted the loss functions employed~\cite{arjovskyPrincipledMethodsTraining2017,maoLeastSquaresGenerative2017,gulrajaniImprovedTrainingWasserstein2017}, progressively scaled generate size~\cite{karrasProgressiveGrowingGANs2018}, or employed style transfer techniques~\cite{karrasStyleBasedGeneratorArchitecture2019,karrasAnalyzingImprovingImage2020} to improve training stability. These techniques have enabled megapixel resolution generation while retaining much of the quality and diversity of reference images. 

We focus on the Progressively Growing GAN (PGGAN) methodology employed by Karras \emph{et al.}~\cite{karrasProgressiveGrowingGANs2018}. This technique trains on progressively larger images as each resolution converges, stabilising the training process and enabling initial training to progress faster as batch sizes can be larger. It is an attractive option for high resolution image synthesis as it has far more moderate computational requirements compared to its later variants~\cite{karrasStyleBasedGeneratorArchitecture2019,karrasAnalyzingImprovingImage2020}. The training methodology for PGGAN additionally combats the effects of generator modal collapse through the inclusion of minibatch discrimination, which operates by considering the variability of samples it is trained on. This promotes the generator to maintain sample diversity as a reduction would aid the discriminator in identifying samples as fake. Despite these benefits, there are important theoretical limitations for the PGGAN that must be considered. The later StyleGAN variant by Karras \emph{et al.}~\cite{karrasStyleBasedGeneratorArchitecture2019} discusses that the network's only source of stochastic variation arises from the initial latent space sample which results in the consumption of network synthetic capacity to preserve variation for later image scales. This results in repetitive patterns in images and a loss of variability at greater resolutions. In addition, the progressively growing technique itself is known to result in phase artifacts with sections of images becoming fixed in preferred locations from training at lower resolutions~\cite{karrasAnalyzingImprovingImage2020}.

\begin{figure}
    \centering
    \includegraphics[height=8cm]{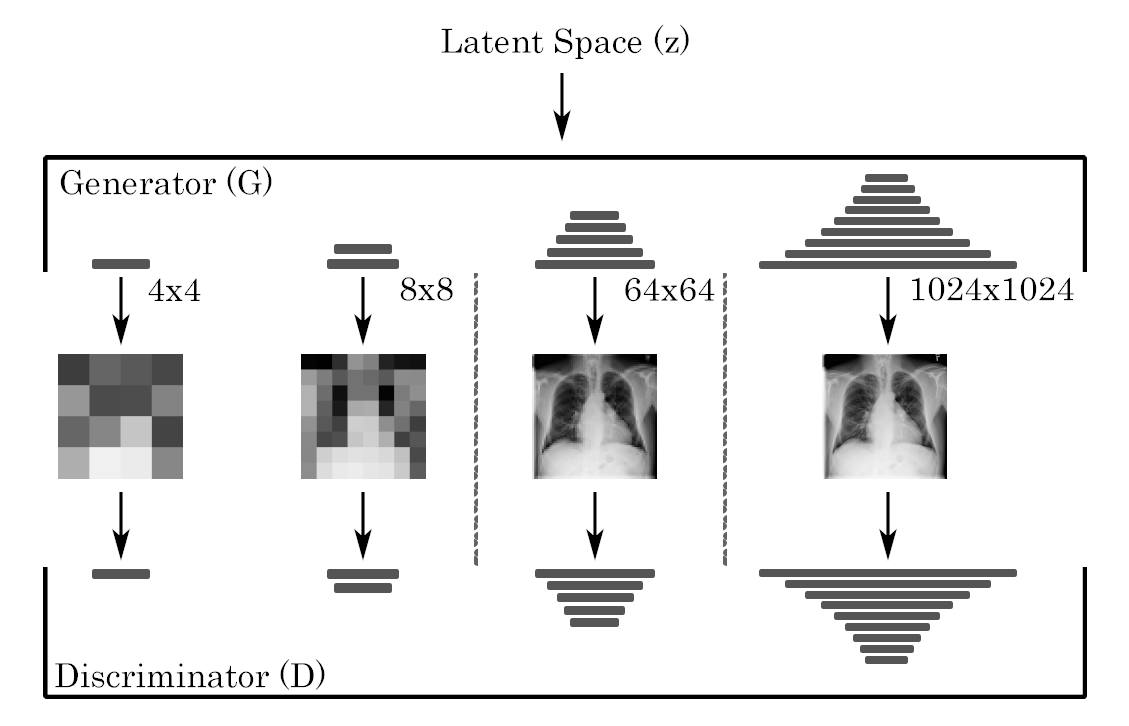}
    \caption{Training configuration for PGGAN. The network trains until convergence and then doubles the spatial resolution. This process is repeated until the desired resolution is achieved.}
    \label{fig:architecture}
\end{figure}

\subsection{Synthetic Medical Data}
There is a multitude of work that aims to adapt GANs for resolving data disparities via data generation while improving both patient confidentiality and model performance. In the pursuit of this, synthetic medical data aims to capture information of diagnostic utility while eliminating the possibility of patient re-identification~\cite{parkDataSynthesisBased2018}. Park \emph{et al.} utilise GANs to produce anonymised clinical data tables that are interoperable with model architectures applied to standard tables, yet can be shared without concern of violating patient privacy~\cite{parkDataSynthesisBased2018}. 

Other works have successfully applied DCGANs for dataset augmentation in various CXR classification tasks. Moradi \emph{et al.} focus on generating both normal and cardiac abnormality images, they find that GAN-based augmentation outperforms traditional training augmentation of flipping, cropping, or scaling images~\cite{moradiChestXrayGeneration2018} . Salehinejad \emph{et al.} expand upon this concept by producing a per-class DCGAN and utilising generates to balance classes for a more diverse performance improvement~\cite{salehinejadSynthesizingChestXRay2019a}. Both approaches demonstrate the potential benefits GANs offer for dataset augmentation, however they are handicapped by the need for per-class models. Training a model per class is computationally prohibitive and fails to capture the interactions between pathologies that may be exploited to improve diagnostic accuracy.

Beyond CXRs, PGGANs have been applied to generate other high resolution medical images. Beers \emph{et al.} demonstrate the applicability of the method for creating high fidelity reproductions of the retinal fundus and MRI slices of gliomas ~\cite{beersHighresolutionMedicalImage2018}, while Togo \emph{et al.} produce patches of x-rays of gastritis~\cite{togoSyntheticGastritisImage2019}. Both implementations introduce domain-specific modifications. Beers \emph{et al.} include segmentation maps as additional channels for fundal images to enhance the generation of features relevant to diagnosis of retinopathy, while Togo \emph{et al.} introduce a conditional loss at higher resolutions to promote the distinction between gastritis and normal tissues.  Korkinof \emph{et al.} synthesise high resolution full-field mammograms and provide comparisons of generated and source images, noting that the images appear similar with the preservation of several common tissue artifacts~\cite{korkinofHighResolutionMammogramSynthesis2019}. Bowles \emph{et al.} augment CT and MRI data with synthetic slices and observe improvements in segmentation performance~ \cite{bowlesGANAugmentationAugmenting2018}.

Despite significant variation in both domain and task, most medical image GAN implementations are constrained to produce individual classes or random samplings from the latent space with little control over generated content.

\subsection{Image Quality Metrics}
The automated evaluation of the quality of medical imagery is a diverse field with a variety of available methods dependent on the evaluation task at hand. These methods are typically divided into full, reduced, and no reference assessment methods. Full reference methods typically evaluate degradation of an image against a source. This may take many forms, from directly comparing pixel values, signal to noise ratios, or to assessing structural or feature similarities. Reduced reference methods evaluate alterations in images against natural image statistics and are typically used in finding distortions in transmitted images. No reference methods only compute elements of a given image to produce a quality assessment and are similarly utilised for transmitted images where no reference may exist. Utilising standard full reference methods for synthetic image evaluation is difficult as while the images may closely resemble the source image distribution, the generator does not have direct access to the distribution and as such does not reproduce any particular image. This precludes the use of standard comparison techniques as even structural or feature similarity metrics are intended for comparison of matched images~\cite{thankiMedicalImagingIts2018}. 

The Fréchet Inception Distance (FID) is a metric intended to provide a solution for evaluating the quality of generated images~~\cite{heuselGANsTrainedTwo2018}. It functions by embedding a set of real and synthetic images in the final average pooling layer of an Inception Net~~\cite{szegedyRethinkingInceptionArchitecture2015} pre-trained on ImageNet~~\cite{russakovskyImageNetLargeScale2015}. The sets are assumed to be multivariate Gaussian distributions with the average and covariance of each utilised to calculate the Fréchet distance (\ref{eq:FID}), also known as the Wasserstein-2 distance. $(\mu_{P_{r}}, \sum_{P_{r}})$ and $(\mu_{P_{g}}, \sum_{P_{g}})$ refer to the mean and covariance of the real and generated distributions respectively.

\begin{equation} \label{eq:FID}
\text{FID}(P_{r}, P_{g}) = \norm{\mu_{P_{r}} - \mu_{P_{g}}}^2_2 + \Tr{(\sum{_{P_{r}}} + \sum{_{P_{g}}} - 2(\sum{_{P_{r}}}\sum{_{P_{g}}})^\frac{1}{2})}
\end{equation}

This distance reflects the difference in the average features extracted from each image set based on the learned kernels of the Inception Net model. This is broadly similar to the feature-based full reference quality assessment metric, yet is capable of application to unpaired images by considering the average combination of features in both sets. The distance has been demonstrated to be consistent with human judgement of visual quality and more resistant to noise than prior approaches~\cite{heuselGANsTrainedTwo2018,lucicAreGANsCreated2018}. The FID is sensitive to class mode dropping, with distances increasing with greater class discrepancies between the two sets as the average set of features begins to differ. Despite this sensitivity, the FID reports ideal results if a model reproduces the training samples perfectly~\cite{lucicAreGANsCreated2018}. The use of the FID for medical images requires the underlying Inception model be capable of extracting features to adequately represent and compare sets against one other. This may be problematic given that x-rays are distinct from the classes found in ImageNet.

Human eYe Perceptual Evaluation (HYPE)~\cite{zhouHYPEBenchmarkHuman2019} in contrast to the aforementioned automated approaches, standardises the human evaluation of model generates by considering either the time necessary to discriminate between real and fake images (HYPE$_{time}$) or the average error rate given unlimited evaluation time (HYPE$_\infty$). The evaluation methodology is demonstrated to produce reliable results that are capable of separating out differences in model performance.

In addition to the already mentioned quality measures for generated images, a significant element to examine is the performance of models developed using synthetic data applied to the original task of interest. Shmelkov \emph{et al.}~\cite{shmelkovHowGoodMy2018} propose the GAN-train metric, which evaluates the classification performance of a model trained exclusively on synthetic data and then run on the original test set. This method quantifies the effective difference between synthetic and real datasets on a benchmark task of interest and provides important information regarding the extent to which the model captures significant features of the underlying data distribution.

\section{X-Ray Synthesis Experiments}
\subsection{Dataset}
\label{sec:dataset}
The ChestX-ray14 dataset is an update to the ChestX-ray8 dataset~\cite{wangChestXray8HospitalscaleChest}, a large, open medical X-Ray dataset published by the National Institutes of Health (NIH) Clinical Center. The dataset comprises \num{112120} x-rays from \num{30805} unique patients. The images comprise \num{15} classes, viz.:
\begin{AutoMultiColItemize}
    \item No Finding
    \item Atelectasis
    \item Cardiomegaly
    \item Consolidation
    \item Edema
    \item Effusion
    \item Emphysema
    \item Fibrosis
    \item Hernia
    \item Infiltration
    \item Mass
    \item Nodule
    \item Pleural Thickening
    \item Pneumonia
    \item Pneumothorax
\end{AutoMultiColItemize}
These images were collected from a clinical archive and should broadly reflect the typical clinical prevalence of these conditions within the community served by the NIH. 75\% of images are normal investigations. The remainder are made up of the various labels ranging from the most prevalent, infiltration (10\%), to the least, hernia (0.5\%). The diagnostic labels are accompanied by bounding boxes for feature localisation. These labels were created through natural language processing (NLP) of the associated radiology and were initially estimated to be over 90\% accurate. This accuracy has been disputed, with the visual content reviewed to not adequately match the proposed labels for a number of investigations~\cite{oakden-raynerExploringLargeScale2019}. A modified set of labels was made available by Rajpurkar \emph{et al.}~\cite{rajpurkarDeepLearningChest2018} as part of their work on pathology detection, wherein a network was trained to classify images based on the original labels and then subsequently used to relabel the original dataset. These labels are available for the majority of images with a residual manually labelled test set which has not been released publicly.

The NIH dataset, license and publication can be found here: \url{https://nihcc.app.box.com/v/ChestXray-NIHCC}

\subsection{Progressively Growing GAN}
\label{sec:gan-train}
We implement a modified PGGAN model~\cite{karrasProgressiveGrowingGANs2018} in Pytorch and perform training via Pytorch Lightning~\cite{falconPyTorchLightning2019} based on the open-source implementation produced by Facebook Research and available through the Pytorch Model Zoo~\footnote{\url{https://github.com/facebookresearch/pytorch_GAN_zoo}}. We start with a randomly initialised model and initially generate 4x4 images with progressively doubling of the spatial resolution after convergence at each scale. We allow mixing of the prior trained layers with the newly added layers through upsampling with the proportion of the previous layer considered reducing in linearly as training progresses.

We utilise WGAN-GP loss~\cite{gulrajaniImprovedTrainingWasserstein2017}, equalised learning rates, minibatch discrimination, and pixel normalisation~\cite{karrasProgressiveGrowingGANs2018}. We also implement an exponential moving average of generator weights which we use for evaluation. We train up to a resolution of 1024x1024 with \num{800000} images shown during each period of layer mixing and a further \num{800000} for training of the added layers. All training was performed on an Amazon Web Services (AWS) p3.8xlarge instance with 4 V100 GPUs. Model training took 6 days.

\begin{figure}
    \centering
    \includegraphics[height=13cm]{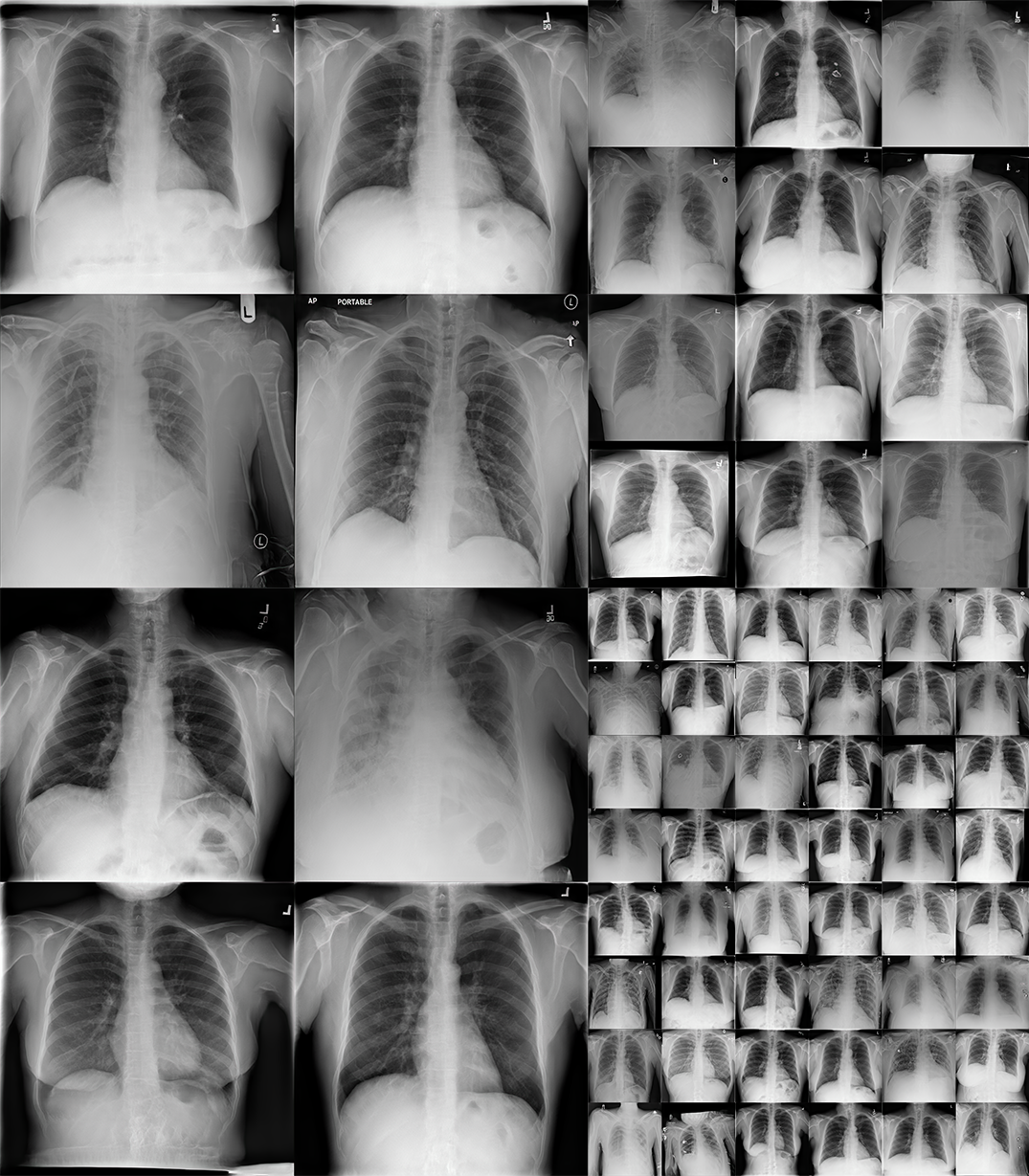}
    \caption{Examples of uncurated, random samples from the exponential moving average of the generator. Samples are drawn at random from a $\mathcal{N}(0, 1)$ distribution without use of the truncation trick.}
    \label{fig:collage}
\end{figure}

Examples of images generated by the model can be seen in figure \ref{fig:collage}.

\subsection{Image Classification}
\label{sec:classifier-train}
For labelling the various CXRs, we implement a Densenet-121~\cite{huangDenselyConnectedConvolutional2018} pre-trained on ImageNet~\cite{russakovskyImageNetLargeScale2015} and replace the final fully connected layer with the number of classes in the ChestX-ray14 dataset. The model is trained end-to-end in a multi-label configuration to predict all classes simultaneously while making use of the weighted cross entropy function (\ref{eq:BCE-weight}) implemented by Guendel \emph{et al.}~\cite{guendelLearningRecognizeAbnormalities2018}. The modified loss balances the frequency of positive ($N_p$) and negative ($N_n$) labels per class ($c$) based on the overall frequency within the training dataset. This adjusts the loss to require the model to discriminate equally between all classes which improves performance for rarer classes.

\begin{equation} \label{eq:BCE-weight}
\begin{split}
D_{CE}(y_c,\ \hat{y}_c) = w_P \cdot y_c \log(\hat{y}_c) + w_N \cdot (1 - y_c) \log(1 - \hat{y}_c),  \\
w_P = \frac{N_p + N_n}{N_p}, \quad
w_N = \frac{N_p + N_n}{N_n}
\end{split}
\end{equation}

For training, we extract a subset of the full dataset to include only images with modified labels from Rajpurkar \emph{et al.}~\cite{rajpurkarDeepLearningChest2018}, we group images at the patient level and average the labels across all images for an individual to produce a summary of the average set of conditions per patient. The dataset is then split into training, validation and test sets through iterative stratification of the average patient labels to maintain label proportions across sets while ensuring no patient overlap occurs. We augment images with a 10\textdegree \ rotation, random horizontal flip probability of 50\%, and colour jitter for brightness, contrast, saturation, and hue of \num{0.1}. The model is trained using an ADAM optimiser~\cite{kingmaAdamMethodStochastic2015} using the default settings with an initial learning rate of $10^{-3}$, which we reduce by a factor of ten if the validation macro-averaged Receiver Operating Characteristic (ROC) Area Under Curve (AUC) fails to improve for several epochs. Once model performance plateaus, it is evaluated on the test set. We repeat the same process for the final trained discriminator, replacing the final linear scoring layer with the number of classes and train the subsequent classification model end-to-end.

\subsection{Pathology Optimisation}
\label{sec:image-optimiser}
Given the multi-modal nature of pathology contained within x-ray images, it is not possible to utilise standard conditional GANs~\cite{mirzaConditionalGenerativeAdversarial2014,odenaConditionalImageSynthesis2017a} for label-specific image synthesis, as these typically require independent classes that can be encoded and fed to the generator to control synthesis. Multi-modal constraints are more consistent with text-to-image generation~\cite{reedGenerativeAdversarialText}, which considers a vector representation of the conditional text input that is semantically meaningful and capable of providing a useful distance based on model predictions~\cite{mirzaConditionalGenerativeAdversarial2014}. This enables the discriminator to evaluate both image quality and similarity to the conditional text input in scoring generates. These techniques are likely to be problematic in the progressively growing formulation as the disease patterns vary considerably in terms of scale of finding and influence over the image as a whole. As an example, \emph{Cardiomegaly} may be evaluated simply by having the cardiac border enlarged relative to the overall width of the chest cavity, whereas \emph{Emphysema} in contrast may have enlarged lung fields, flattened diaphragms, and a reduced apparent cardiac size. These features become apparent at differing resolutions and as such, attempting to condition this information throughout training is likely to hamper performance. 

\begin{figure}[!htb]
    \centering
    \includegraphics[height=8cm]{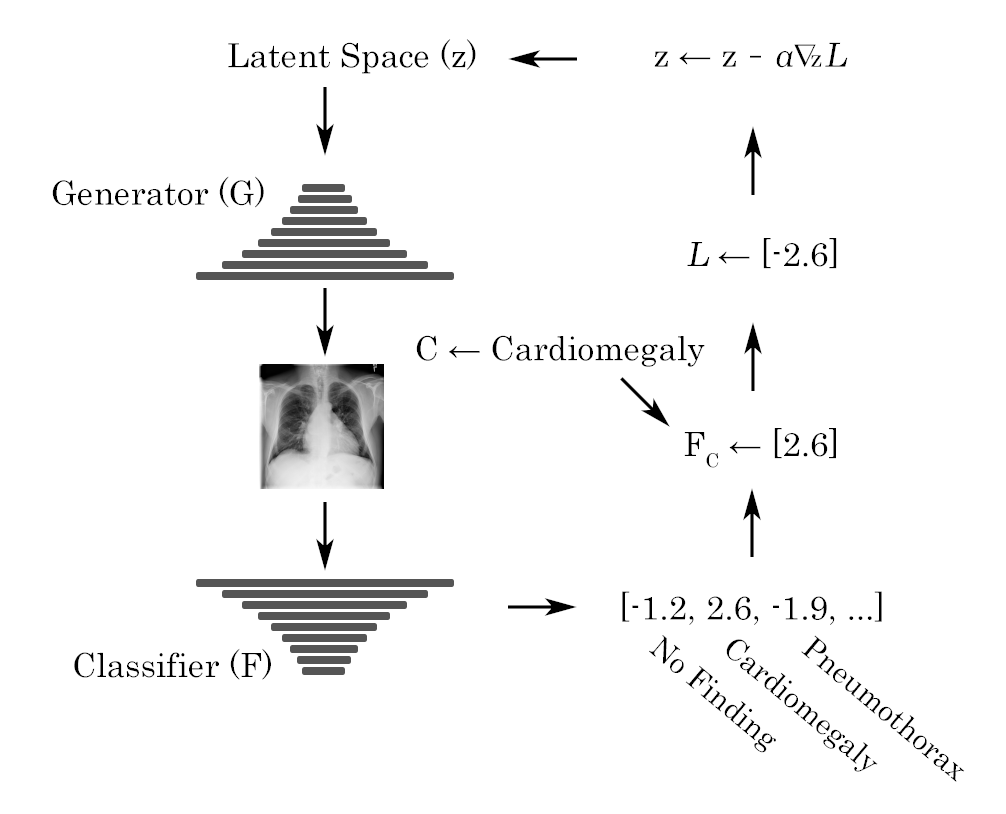}
    \caption{Visual representation of class generation method shown in Algorithm \ref{opt-algo}. }
    \label{fig:class-optim}
\end{figure}

We sidestep these issues in multi-modal conditional training by instead drawing inspiration from work on the embedding of images into the latent space~\cite{creswellInvertingGeneratorGenerative2016}. Embedding images requires the inversion of the generator. This is typically performed either by training an encoder network that maps an image to a location in the latent space or by gradient descent optimisation of a random sample to minimise reconstruction loss~\cite{abdalImage2StyleGANHowEmbed2019}. We opt to find a latent representation that maximises the corresponding image's classification score for a label of interest. We optimise using a single label at a time to allow for inclusion of related labels that would be consistent with the presentation of real x-rays. We base the implementation on the work by Creswell \emph{et al.}~\cite{creswellInvertingGeneratorGenerative2016}. Pseudocode for such a method can be seen in Algorithm \ref{opt-algo} with a visual representation of this technique in figure \ref{fig:class-optim}. 

\begin{algorithm}[H]
\label{opt-algo}
\SetAlgoLined
$z^* \sim P_z(Z)$ \tcp*{Initialise $z$ from prior distribution}
$S \leftarrow F_c(G(z^*))$\;
 \While{$S < T_c$}{
 $x \leftarrow G(z^*)$ \tcp*{Generate image proposal}
 $S  \leftarrow  F_c(x)$ \tcp*{Obtain class score from classifier}
 \If{$S \geq T_c$}{
 break\;
 }
 \Else{
 $L  \leftarrow  -S$\;
 $z^*  \leftarrow  z^* - \alpha  \nabla_z L $ \tcp*{Apply gradient descent}
 }
 }
 \Return{$z^*$}
 \caption{Algorithm for inferring $z^* \in Z$ , a latent representation for a generator $G$ that produces an image $x \in R^{m \times n}$ representative of class $c \in C$ determined by a classifier $F$ exceeding a class specific threshold $T_c$}
\end{algorithm}

\section{Results \& Analysis}
\label{sec:analysis}
\subsection{Image Quality}
\label{sec:img_quality}
The generated images, as seen in figure~\ref{fig:collage}, appear to broadly reflect the NIH source images with global features varying similarly across both sets. Sex, posture, exposure, and positioning (AP vs PA) are all represented within the generates. The images accurately reflect the standard anatomical features found in CXRs with realistic alterations in perspective seen with changes in patient posture or positioning of the x-ray detector. Soft tissues such as the heart, liver, and stomach are faithfully reproduced with normal variability in their relative positioning. Bony structures are correctly placed but suffer from inconsistent profiles, with the ribs in particular tending to reveal a degree of undulation. Closer inspection often reveals slight curves or alterations in calibre that are seldom explained by the perspective of the image. Beyond the x-ray itself, the model has learnt to include various markup elements included in the reference images. Most images have a symbol or tag demonstrating the left side of the image, in addition, projection descriptors such as `PORTABLE' and `AP' are included on numerous images. Images are typically sided correctly and tend to have visual projections similar to included labels, although at times this is difficult to confirm. Some images do not match the included text, with some having multiple copies of a particular label. A white arrow typically used to demonstrate that a patient was upright for portable scan has also been reproduced, tends to co-occur with text denoting the scan type as portable and is often in similar positions to the reference images. Elements such as white borders, poor exposure, cropping, or rotation are all present in the source dataset and reproduced in some samples by the PGGAN model.

\begin{figure}
    \centering
    \includegraphics[height=7cm]{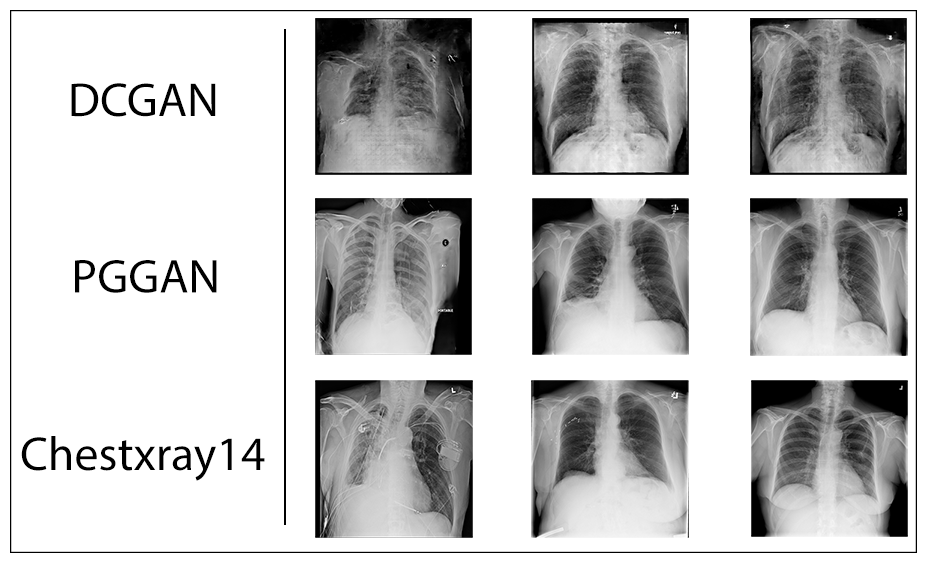}
    \caption{Comparison of sample quality between different GAN architectures and the source dataset.}
    \label{fig:DCGAN}
\end{figure}

The main distinction between distributions is that the generated images broadly lack certain smaller scale features of the source set. Jewelry, ECG leads, pacemakers, and IV catheters are largely absent with occasional partially formed objects in locations where these should appear. The elements of the missing structures form only a fraction of each overall image and are often fairly detailed objects themselves. A potential explanation for this phenomenon is the progressive growing technique itself, as these objects would only be generated near the end of the training process as the resolution of the images allowed for the details to be appreciated. Examining the generator in the final model without the moving weight average, supports this as a larger proportion of the images possess partially formed objects, implying the network was slowly incorporating these features. The appearance of these elements failed to improve despite training beyond the recommended total number of images. If this is the case, it is probably due to the reduced batch size at higher resolutions which would result in exceedingly slow convergence for rarer, more subtle features. An alternate explanation is the aforementioned capacity loss that prompted the shift to style-based networks as networks consume their own synthetic ability to retain variability in later layers. The effects of this are likely to be more pronounced in CXRs compared to other domains, like facial data, due to the presence of distinct small scale structures such as IV lines or ECG leads that are readily discernible on standard images. 

To evaluate the applicability of the FID for automatic x-ray quality assessment, we first attempt to produce a meaningful zero measure by splitting the NIH dataset at the patient level and using the average labels from each patient to iteratively stratify the patients into two similar groups. The concept being that the groups should have similar disease distribution and be free of patient overlap. We calculate a low value of \num{0.53} for the FID, indicating that the underlying Inception network extracts similar features on average from both sets. These features may be sub-optimal given that the underlying network has not been trained on biomedical data, yet the small distance demonstrates that the images are embedded in a similar manner. We now need to demonstrate whether the metric can in fact distinguish between x-rays.

Given that the images are known to vary by pathology label, we argue that if the distance metric can reliably separate distinct classes, it must have filters that are capable of extracting features that are meaningful for evaluating the clinical plausibility of generates. If we were to re-train an Inception network purely to calculate the FID for medical imagery, the network would be trained to classify an image according to its class labels. If the current weights of the network are already able to provide such a separation, then there is no need to re-train and the metric can in fact be applied to the problem at hand. 

\begin{figure}[!htb]
    \begin{subfigure}{0.5\textwidth}
    \includegraphics[width=8cm]{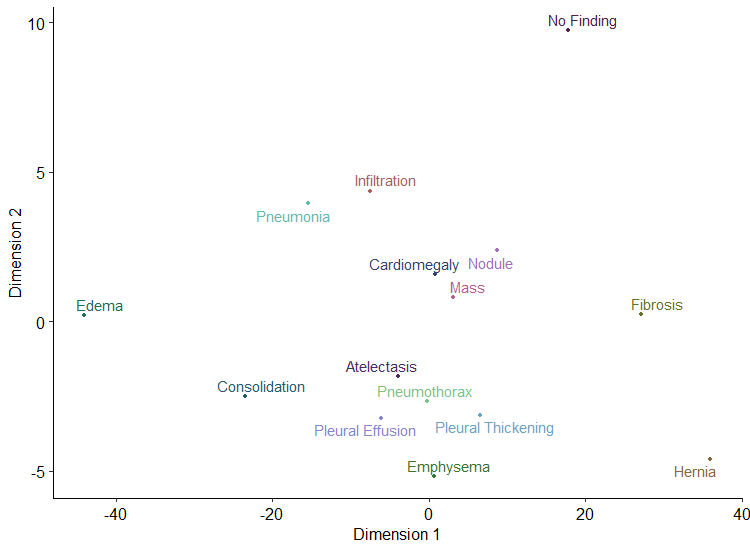}
    \end{subfigure}
    \begin{subfigure}{0.\textwidth}
    \includegraphics[width=8cm]{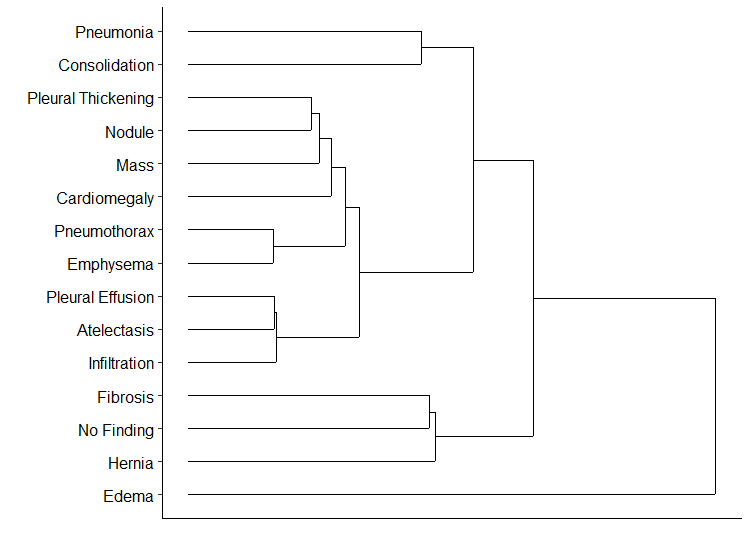}
    \end{subfigure}
    
    \caption{NIH Class Label FIDs. Left: Classes plotted according to the multidimensional scaling of the pairwise distance matrix calculated by selecting each class as a baseline and determining the distance to all other classes in turn. Right: A hierarchical cluster based on the distance matrix showing the grouping of similar classes based on the features extracted by the Inception network when calculating the FID.}
    \label{fig:class-dist}
\end{figure}

To evaluate the extent to which the FID can discriminate between individual pathologies we utilise the \emph{No Finding} label as a baseline as it principally should be absent of any significant changes while still varying similarly along protected patient parameters such as sex and age. We then evaluate the distance between it and the set of x-rays containing each other label, the results of which can be seen in table with the addition of a split along sex and the overall comparison between real and synthetic images~\ref{table:FIDs}. The results provide evidence that the FID is able to distinguish between the labelled pathologies and a baseline x-ray with the distance seemingly recapitulating the relative scale of the pathological change on the image. Very large distances are associated with the \emph{Edema} and \emph{Consolidation} labels, which are findings that may distort major segments of the lung fields, while \emph{Mass} and \emph{Nodule} labels produce significantly reduced distances as they generally occupy only small segments of the study. To provide more substantial evidence,  we calculate a pairwise distance matrix between all classes and utilise multidimensional scaling to plot the relative 2D positions in figure \ref{fig:class-dist}. The various class locations show that the FID is able to cluster similar labels together. In addition, a hierarchical cluster based on the distance matrix, also seen in figure \ref{fig:class-dist}, successfully groups sets of related findings. The cluster demonstrates that the features extracted by the network underlying the FID are likely to be of a high enough quality to group sets of related findings and that the technique is likely applicable to CXRs despite not being explicitly trained to detect features of these images.

\begin{table}[ht]
\centering % used for centering table
\caption{Fréchet Inception Distance (FID) per Dataset Split} % title of Table
\begin{tabular}{l c l c} % centered columns (4 columns)
\hline\hline %inserts double horizontal lines
Split & FID & Split & FID\\ [0.5ex] % inserts table
%heading
\hline % inserts single horizontal line
Stratified & 0.53 & Hernia & 15.96 \\
Sex & 7.87 & Infiltration & 20.05 \\
No Finding & 0.00 & Mass & 10.06 \\
Atelectasis & 19.90 & Nodule & 6.22 \\
Cardiomegaly & 14.23 & Pleural Effusion & 23.90 \\
Consolidation & 42.45 & Pleural Thickening & 13.05 \\
Edema & 59.40 & Pneumonia & 32.05 \\
Emphysema & 19.56 & Pneumothorax & 18.00 \\
Fibrosis & 9.72 & Synthetic & 8.02\\ [1ex] % [1ex] adds vertical space
\hline %inserts single line
\end{tabular}
\label{table:FIDs} 
\end{table}

We as a result evaluate the quality of synthetic x-rays compared to the NIH source images. The FID between the full source dataset and \num{100000} random PGGAN generates is \num{8.02}. This value is comparable with FIDs reported for the PGGAN architecture on other non-biomedical high resolution image generation tasks \cite{karrasProgressiveGrowingGANs2018}.

\subsection{Pathological Variability }
\label{sec:variability}
\begin{figure}[!htb]
    \centering
    \includegraphics[width=8cm]{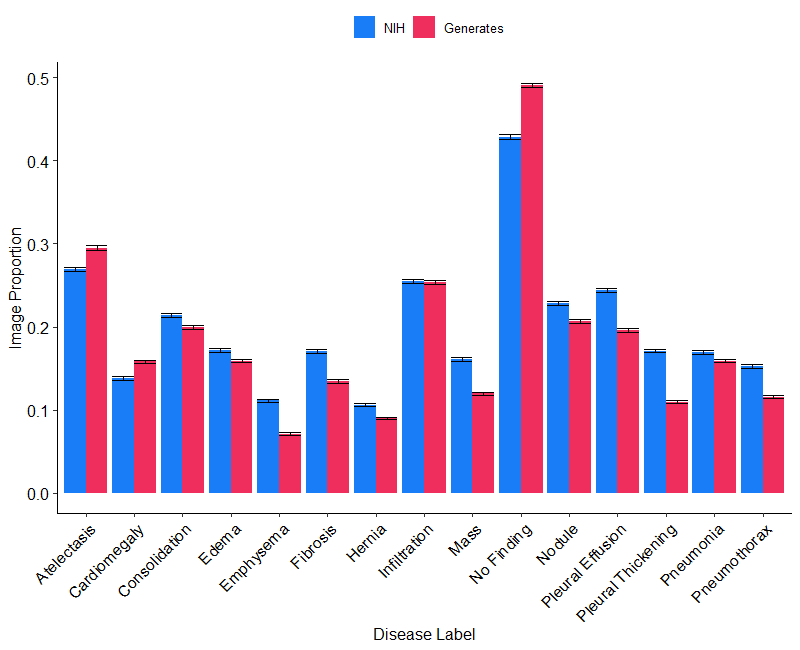}
    \caption{Label prevalence of the original NIH dataset and a random selection of \num{130000} PGGAN generates. Point estimates are the average number of each label across all images. 95\% Confidence intervals are provided by re-sampling the full set of labels with replacement \num{10000} times. The network demonstrates a degree of mode dropping of classes with finer details that it finds difficult to reproduce.}
    \label{fig:prevalence}
\end{figure}

We estimate the pathological variability of the PGGAN generates in an effort to quantify the extent to which model capture the variability of the various disease entitites contained within the source CXRs. 
The labels were estimated by utilising a common classifier to label both the NIH and generated iamge sets with confidence intervals derived by bootstrapping the predictions \num{10000} times. The proportion of the class labels from the NIH dataset and a random selection of \num{130000} generated images can be seen in Figure \ref{fig:prevalence}. Overall it can be seen that the generates cover the range of labels with each represented in broadly similar proportions to the NIH set. Despite the similarity, there appears to be a degree of mode dropping with a majority of labels being significantly less prevalent in the generates compared to the reference set with the exception of \emph{Atelectasis}, \emph{Cardiomegaly}, \emph{Infiltration} and \emph{No Finding}. We hypothesize that this is likely to be due to a continuation of the phenomenon noted with the absence of certain smaller objects, the objects in this case being features of disease. This is supported by the label distribution, as more common conditions with larger disease features tend to be over represented, while diseases with progressively finer features are increasingly sparse in the generated samples. \emph{No Finding} for example, is largely defined by the absence of smaller features and can be seen to be significantly in excess beyond the NIH set. \emph{Atelectasis} and \emph{Cardiomegaly} in comparison, define several characteristics that are present at a moderate resolution and see a smaller increase in prevalence. This is in stark contrast to \emph{Emphysema} and \emph{Pleural Thickening}, which may be quite fine on x-ray and similarly show the greatest reduction in prevalence. 

\subsection{Pathology Generation}
\label{sec:discriminator}
\begin{figure}[!htb]
    \centering
    \includegraphics[height=13cm]{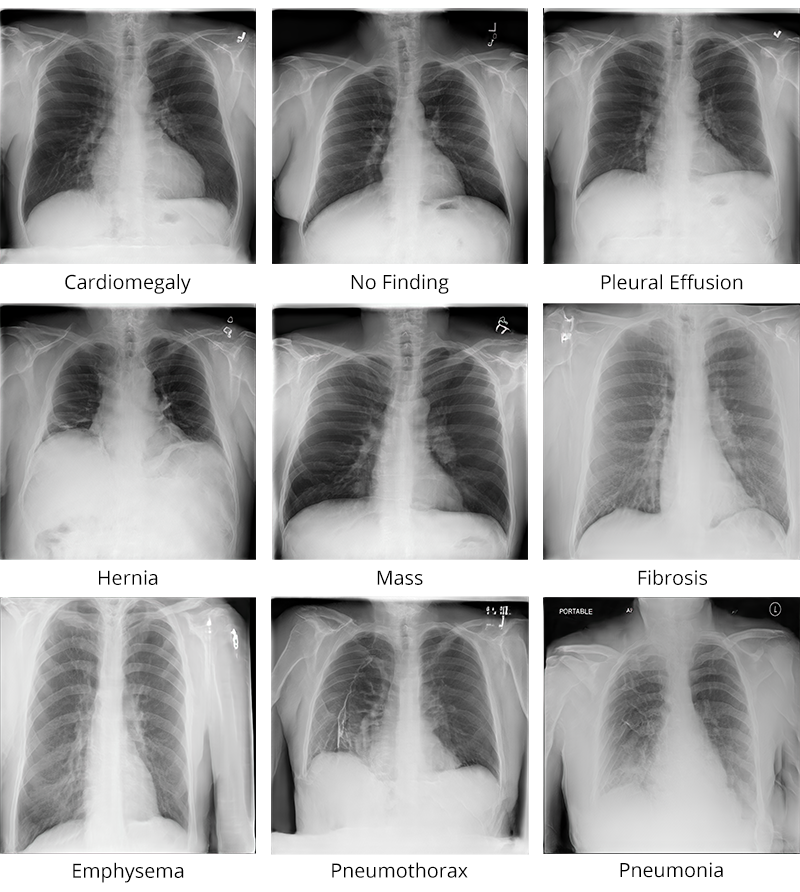}
    \caption{Example classes generated by maximising the classifier class logit. Images often have multiple findings, the finding optimised is the label given.}
    \label{fig:class-pos}
\end{figure}
We utilise class optimisation to generate pathological images based on the method described in Algorithm \ref{opt-algo}. We sample the initial latent code from a truncated normal distribution with a threshold of \num{0.7}~\cite{brockLargeScaleGAN2019a}. We attempt the method with both a Densenet-121 and a re-purposed Discriminator as classifiers. We generally found optimising the Densenet-121 model proved to be slower, more difficult and less reliable overall. The process frequently plateaus and subsequently fails to produce a sample for a particular class. Successes often rely on favourable sampling from the latent space, with such samples typically converging rapidly. The same process applied to the re-purposed discriminator proved more successful as samples converged more regularly with greater ease in achieving higher logit values for a particular class. This discrepancy is due to the input size of the different models. The Densenet as a result of being trained on a smaller resolution requires images to be resized prior to classification and as such results in a worse estimation of the gradient. Replication of the process with a Densenet trained on a larger input size produced significant improvements to both the reliability and rapidity of convergence. Examples of successful pathology generates can be seen in Figure~\ref{fig:class-pos}. Many more examples of multi-modal pathology can be seen in the linked image archive. 

Attempts to modify the optimisation process to produce examples of isolated labels often fails to converge, with optimisation by minimising other class scores alongside the maximisation of the class of interest typically resulting in a significant limitation to the score that can be achieved. This phenomenon is explained by the multi-modal nature of the disease classes and the underlying biomedical relationship between classes that results in label co-occurrence. These relationships can be seen with increases to the logit threshold when optimising for a particular class. Clearer examples of particular conditions tend to co-occur with their medical complications. Increasing class scores for \emph{Cardiomegaly} begin to include a greater proportions of \emph{Pleural Effusion} labels, this may be related to the complications that occur in heart failure that may cause images with these labels to co-occur. Similarly, increasing \emph{Emphysema} class values produces associated \emph{Pneumothorax} labels, while \emph{Nodule} findings produce associated \emph{Mass} labels. These findings reflect that the underlying unsupervised training methodology has captured the relationship between various x-ray features and optimising for the pathology enables us to tease out these properties.

As part of class optimisation, we're required to set a score threshold for determining when a proposed image has converged to be representative of a particular class. We utilise Youden's J statistic to determine the optimum cut-off based on the validation ROC curve per class when training the classifier. To simulate degrees of disease severity in produced images, we randomly increase this threshold by the absolute value of the distribution of $\mathcal{N}(1, 1)$ when optimising. It is quite seldom that a dataset includes only a single image of a patient. Typically there are series of patient images where individuals are re-scanned several times. There are a multitude of potential reasons for this, follow-up on patient condition, assessment of medical intervention, or routine imagery prior to surgery, are just a handful of potential reasons. The end result is that most datasets have a significant proportion of very similar images wherein position or disease severity may be altered slightly but the overall image is largely unchanged. We simulate this process by sampling in the vicinity of a particular pathology-optimised latent code. We treat the optimised location as a centre and sample the surrounding hypersphere within $\mathcal{N}(0, 0.2)$. We find that this enables the creation of images with the same set of diagnoses but minor changes in orientation, exposure, or disease severity. The degree of variability can be tuned by adjusting the standard deviation, however, we find a value of \num{0.2} tends to produce minor variations without compromising the diagnostic label.

\begin{figure}
    \centering
    \includegraphics[height=3.5cm]{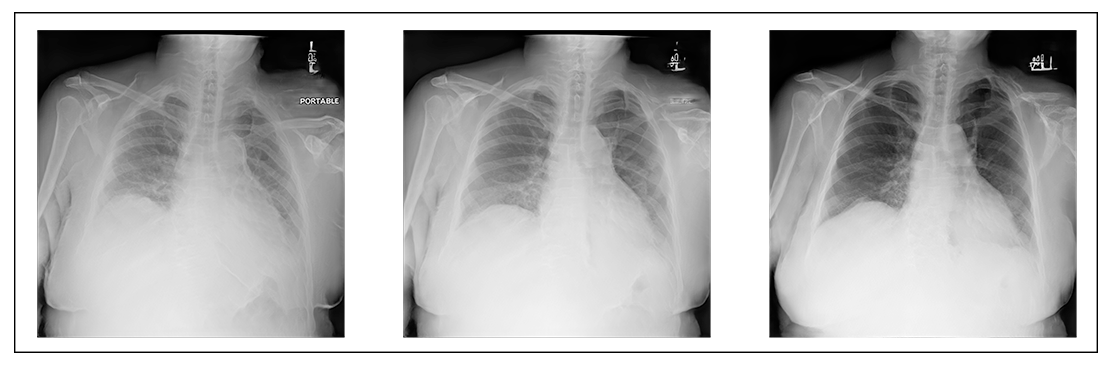}
    \caption{Example of local sampling after class generation. Images show similar anatomical and pathological features with variations in image contrast, text markup, and patient positioning.}
    \label{fig:local}
\end{figure}

The fact that the latent space may be optimised with regard to disease classification at all implies a degree of orientation with respect to pathology which may hold semantic value. We leave a full exploration of the properties of the latent space to future work.

\subsection{Radiologist Review}
\label{sec:review}
We evaluate the plausibility of generated x-rays to domain experts through a series of image reviews. We ask a group of practising radiologists to identify a set of real images out of a grid of mixed x-rays as well as to select a real image out of a pair of real and generated images. We received feedback from 3 consultant (equivalent to board certified) radiologists, and 5 radiology registrars (equivalent to residents in North America). Image grids consist of six x-rays that are produced with a \num{50}\% probability of being real or generated, we evaluate six such grids. Real images were identified as such by radiologists \num{73}\% (95\% CI: \num{63}, \num{82}) of the time, while generates were identified as real \num{61}\% (95\% CI: \num{51}, \num{70}) of the time, with both groups more likely than chance to be identified as real, with reals p < \num{0.0001} and generates p = \num{0.0155} for a one-sided t-test. 
Radiologist performance on x-ray discrimination equates to a HYPE$_\infty$ score of \num{33}\% (95\% CI: \num{24}, \num{43}).  
Discrimination of image pairs is clearly in favour of real images, with radiologists correctly identifying the true image of a pair \num{71}\% (95\% CI: \num{55}, \num{86}) of the time. We evaluate \num{20} such image pairs. Comments from respondents mentioned bony abnormalities as the main feature used to identify samples as being synthetic. Images were embedded as options within a Google Form and shown at a 260x260 resolution with unlimited time to review images prior to making a decision. 

\subsection{Synthetic Model Training}
\label{sec:synth-model}
For evaluating the classification performance of a model trained only on synthetic data, we generate an archive of CXRs of comparable size and with labels in a similar proportion to the source NIH dataset. We uniformly sample between \num{3} to \num{6} related images per class optimisation operation and treat them as belonging to the same 'patient'. We similarly iteratively stratify the synthetic patients into training and validation sets in the same manner as during initial classifier training, retaining only the original test set of real images, which we use for all comparisons. We train a Densenet classifier in the same configuration as described in section \ref{sec:classifier-train}. We evaluate performance relative to real images as well as images embedded into the generator's latent space. We consider embedding images as the current alternative to individual class generation, whereby a network predicts the latent encoding of a given image. To achieve this, we retrain the discriminator to predict such a representation given a random sample from the generator and minimising the mean squared error between the predicted and actual locations in the latent space. We train with the proposed embedded image alongside the original labels. 
A table of comparative ROC AUC values can be seen in table \ref{table:Synth-AUC}

\begin{table}[ht]
\centering % used for centering table
\caption{Comparison of Performance of Classification Models Trained on Synthetic vs Real Data} 
\begin{tabular}{l c c c} 
\hline 
Class Label & Embedded & Class Generated & Real\\ [0.5ex] 
\hline \hline 
Atelectasis & 0.646 & 0.851 & 0.923 \\
Cardiomegaly & 0.602 & 0.907 & 0.968 \\
Consolidation & 0.675 & 0.928 & 0.963 \\
Edema & 0.746 & 0.941 & 0.985 \\
Emphysema & 0.51 & 0.901 & 0.972 \\
Fibrosis & 0.63 & 0.868 & 0.961 \\
Hernia & 0.611 & 0.951 & 0.986 \\
Infiltration & 0.596 & 0.796 & 0.847 \\
Mass & 0.516 & 0.825 & 0.936 \\
No Finding & 0.605 & 0.861 & 0.932 \\
Nodule & 0.543 & 0.748 & 0.909 \\
Pleural Effusion & 0.636 & 0.903 & 0.954 \\
Pleural Thickening & 0.583 & 0.881 & 0.954 \\
Pneumonia & 0.595 & 0.932 & 0.975 \\
Pneumothorax & 0.568 & 0.877 & 0.947 \\
\hline %inserts single line
Average & 0.604 & 0.878 & 0.947 \\
\hline %inserts single line
\end{tabular}
\label{table:Synth-AUC} 
\end{table}

The class generation results provide further evidence that the PGGAN model has learnt to replicate all the classes within the original NIH dataset, as the synthetic classifier is able to detect each pathology from the source to a degree significantly greater than expected by chance, despite never actually seeing an image from the original dataset. Furthermore, the generated classes demonstrate that the method reliably produces examples of each class, with a performance reduction similar to that seen with previous PGGAN generation tasks~\cite{togoSyntheticGastritisImage2019}. These results should scale with improvements made to the quality of generated images, as augments to the underlying GAN will reduce the distinction between the synthetic and real sets, which alongside class generation, will allow for even better representations for training. The poor performance of embedded samples are indicative of such a method failing to preserve class-specific information, and show it to be less suitable of a method for producing anonymised clinical archives. A potential alternative to predictive embedding would be to optimise the latent space to minimise image reconstruction loss. Our experiments with this showed promise for more faithful representations, however, the performance was profoundly worse than alternatives, requiring days to convert an archive compared to only a handful of hours for competing approaches. We attempted comparisons with generating classes through a simpler DCGAN architecture similarly trained on the full NIH dataset. We found the generation process intractable as the model often failed to produce class scores sufficient to be considered examples of that class despite extensive optimisation. We attribute this to a combination of mode dropping and lower image quality which precludes the formation of certain classes.

These results provide substantial evidence that class optimisation is effective at producing images that are representative of a particular disease label and can enable a single unsupervised GAN to produce a fully labelled cohort of classes from an anonymised dataset. 

\section{Discussion}
\label{sec:discussion}
Through both automated and expert review of the medical plausibility of generated x-rays as well an evaluation of synthetic model performance we have outlined the properties of the PGGAN architecture applied to CXRs. Expert review demonstrated global coherence of the imagery with reproduction of numerous features associated with varying disease classes. Deficiencies have been noted with a broad absence of detailed, small scale features that are typically easily discernible on x-ray films. Automated comparisons of image features by means of the FID provide evidence that the quality of generates is of a similar standard to typical high resolution tasks where PGGAN is applied.

These limitations are likely explained by stochastic bottlenecks and slower convergence of novel features at higher resolutions and are inherent to the design and training formulation employed by the baseline PGGAN methodology. Future work should follow improvements made to subsequent GAN architectures in the PGGAN lineage~\cite{karrasStyleBasedGeneratorArchitecture2019,karrasAnalyzingImprovingImage2020}, or alternatively examine alternate approaches such as Very Deep Variational Autoencoders (VAEs)~\cite{childVeryDeepVAEs2020} which have similarly shown promise for high resolution image synthesis. Irrespective of the exact mechanism used to generate images, the implementation of multi-modal class optimisation allows for the extraction of disease-representative classes from the latent space that can be utilised to construct labelled synthetic datasets of arbitrary size. 

We suspect this method will prove to be of even greater utility with GAN architectures that possess less entangled latent spaces and could allow for disease optimisation at varying resolutions by considering scale-specific elements of the extended latent space seen in StyleGAN and later works~\cite{karrasStyleBasedGeneratorArchitecture2019}. We envision that such techniques should allow significant control over the severity of disease present in generates and may provide profound capabilities for dataset augmentation. This would allow for applications such as demonstrating potential progression or resolution of disease or retaining disease presentation but shifting patient particular parameters such as age or sex. Scale specific modifications may enable the synthesis of support devices such as IV lines that may be differed against the source image and used for augmentation in segmentation tasks. We envision that continued development along this route will enable task-specific image extraction from broad image generators, as unsupervised GANs trained on large corpora of diverse images should be able to extract a multitude of classes based on the capabilities of the associated classifier, even if such classes were not labelled in the original individual datasets. 

\section{Conclusion}
\label{sec:conclusion}
We have applied a progressively growing GAN model to the task of synthesising high-resolution chest x-ray images for the evaluation of their suitability as a replacement for standard images for the tasks of model development and student education. The overarching goal of this investigation being to improve the protection of patient privacy without compromising data availability. We evaluate the applicability of the Fréchet Inception Distance to the evaluation of synthetic chest x-rays and find that the underlying network is capable of providing a meaningful metric for generate quality despite the difference in data distribution. We demonstrate that it is possible to produce realistic, clinically plausible images that capture much of the variation in standard x-rays, however, there remains a significant need for improvement in the reproduction of small-scale details to achieve truly indistinguishable samples. We demonstrate that the model is capable of reproducing all abnormalities of interest and in similar proportions to the source image distribution despite differences in small scale features. We describe a methodology for the extraction of class representative images from the generator's latent space by optimisation of a classifier score and demonstrate that such a method is capable of constructing a fully labelled synthetic dataset from an unsupervised generator. We describe potential avenues of improvement for generate quality and anticipate that such improvements will enable the production of high quality teaching and training images without the concern of breaching patient confidentiality. We make the source code, final model weights and a large archive of labelled generates used in this study available to the broader research community.
\newline \newline
\textbf{Compliance with Ethical Standards}
\newline
\textbf{Funding} The authors declare that they received no funding for this work.
\newline
\textbf{Conflicts of interest} The authors declare that they have no conflict of interest.
\newline
\textbf{Ethics Approval} An ethics waiver was issued for this work by the Human Research Ethics Committee (Medical) of the University of the Witwatersrand, Johannesburg on 11/08/2020.

% \newpage
\printbibliography

\end{document}